\documentclass[amsmath,amssymb,aps,superscriptaddress,twocolumn,floatfix]{revtex4-2}

\usepackage{graphicx}
\usepackage{caption}
\DeclareCaptionLabelSeparator{dot}{. }
\makeatletter
\def\justified{
	\let\\\@normalcr
	\@rightskip\z@skip \rightskip\@rightskip
	\leftskip\z@skip
	\parindent 0em\relax
	\setlength{\parfillskip}{0pt plus 1fil}}
\DeclareCaptionJustification{justified}{\justified}
\usepackage{subcaption}
\usepackage{amsmath}
\usepackage{braket}
\usepackage{units}
\usepackage{ragged2e}
\usepackage[colorlinks,urlcolor=blue ,citecolor=blue ,linkcolor=blue]{hyperref}
\usepackage{xcolor}
\usepackage{nicefrac} 
\usepackage{lipsum}
\usepackage{nameref}
\usepackage{hyperref}
\usepackage{textcomp} 
\usepackage{urwchancal}
\usepackage{upgreek} 
\usepackage{amsmath} 
\usepackage{here}
\usepackage[shortlabels]{enumitem}

\newcommand{\edd}{\epsilon_\text{dd}}

\begin{document}

\title{Vortices in dipolar Bose-Einstein condensates}

\author{Thomas Bland}
\affiliation{Institut f\"ur Experimentalphysik, Universit\"at Innsbruck, Technikerstra{\ss}e 25, 6020 Innsbruck, Austria}

\author{Giacomo Lamporesi}
\affiliation{Pitaevskii BEC Center, CNR-INO and Dipartimento di Fisica, Universit\`a di Trento, I-38123, Trento, Italy}

\author{Manfred J. Mark}
\affiliation{Institut f\"ur Experimentalphysik, Universit\"at Innsbruck, Technikerstra{\ss}e 25, 6020 Innsbruck, Austria}
\affiliation{Institut f\"ur Quantenoptik und Quanteninformation, \"Osterreichische Akademie der Wissenschaften, Technikerstra{\ss}e 21a, 6020 Innsbruck, Austria}

\author{Francesca Ferlaino}
\email[Correspondence and requests for materials
should be addressed to: ]{francesca.ferlaino@uibk.ac.at}
\affiliation{Institut f\"ur Experimentalphysik, Universit\"at Innsbruck, Technikerstra{\ss}e 25, 6020 Innsbruck, Austria}
\affiliation{Institut f\"ur Quantenoptik und Quanteninformation, \"Osterreichische Akademie der Wissenschaften, Technikerstra{\ss}e 21a, 6020 Innsbruck, Austria}

\begin{abstract}
    Quantized vortices are the hallmark of superfluidity, and are often sought out as the first observable feature in new superfluid systems. Following the recent experimental observation of vortices in Bose-Einstein condensates comprised of atoms with inherent long-range dipole-dipole interactions [Nat. Phys.~{\bf 18}, 1453-1458 (2022)], we thoroughly investigate vortex properties in the three-dimensional dominantly dipolar regime, where beyond-mean-field effects are crucial for stability, and investigate the interplay between trap geometry and magnetic field tilt angle.
\end{abstract}

\date{\today}

\maketitle

\section{Quantum vortices in ultracold gases}
\label{section:methods}

Every time a fluid starts to rotate, a vortex can form. This phenomenon is a universal property of fluid dynamics, and has been observed in many different systems on all possible lengthscales, from the spiral motion of galaxies, to the tornadoes in water sinks or in the atmosphere, from superconductors immersed in a magnetic field to rotating superfluids, such as Helium or quantum gases.
The nature of the fluid strongly affects the motion of the particles and the properties of the supported vortices.
A special case is, in fact, represented by superfluids, which are characterized by the absence of viscosity and are described by a continuous, single-valued wavefunction. The latter feature leads to the quantization of circulation and to quantum vortices.

First theoretically predicted by Onsager and Feynman at the second half of the last century \cite{Onsager1949,Feynman1955}, quantum vortices were indirectly seen in superfluid liquid helium \cite{Careri1961cdv,Rayfield1963eft,Packard1972oon,Yarmchuk1979oos}. The advent of experiments on dilute, weakly interacting quantum gases opened the door to an intense series of direct observations of quantum vortices, providing proof of superfluidity in ultracold gases. The existence of a quantum vortex in a superfluid requires it to accumulate a 2$\pi$ phase winding around a line if the fluid is three-dimensional (3D) or around a point if it is confined in a two-dimensional (2D) geometry. 
Typically, the gas density vanishes approaching the vortex core following a $1/r$-law on a characteristic lengthscale given by the local healing length. 
While the ground state of a superfluid is described by a uniform phase, leading to the absence of superfluid motion, any phase-fluctuating excited configuration is accompanied by a local superfluid motion, according to the relation $\mathbf{v_s}=\frac{\hbar}{m}\nabla \phi$.

Phase curls, and consequently vortices, can be generated in a superfluid in many ways. The seminal experimental work of E. Cornell and co-workers at JILA \cite{Matthews1999via}, J. Dalibard and co-workers at LKB \cite{Madison2000vfi}, and that of W. Ketterle and co-workers at MIT \cite{Abo-Shaeer2001oov} observed the first vortices in BECs, opening up the field of quantum vortices in ultracold atoms. Since then, many experimental techniques have been developed, as illustrated in Fig.~\ref{fig:methods}. In a nutshell, such techniques can be divided into the following categories:

\begin{figure*}[t]
    \includegraphics[width=2\columnwidth]{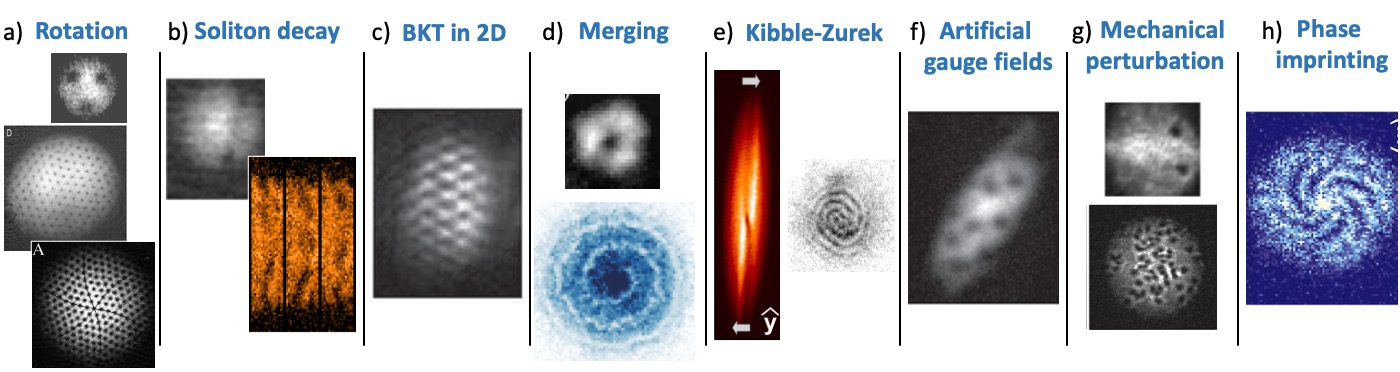}
    \caption{
    Examples of vortices observed in ultracold gases and created with different methods, as listed in Section \ref{section:methods}.
a) Three-vortex sample \cite{Madison2000vfi} and large vortex lattice  (from \cite{Abo-Shaeer2001oov}, reprinted with permission from AAAS) obtained by stirring the BEC with optical potentials at different angular velocities. Vortex lattice obtained by rotating the magnetic trap and showing Tkachenko waves \cite{Coddington2003oot}.
b) Images of topological defects resulting from the decay of an imprinted planar soliton \cite{Anderson2001wds,Ku2016cos}. 
c) Interference between expanding 2D systems containing thermally activated vortices \cite{Hadzibabic2006bkt} near the BKT transition.
d) Interference pattern between three \cite{Scherer2007vfb} or many \cite{Aidelsburger2017rdi} BECs on a ring.
e) Signature of a solitonic vortex in an elongated gas \cite{Donadello2014oos} and interference pattern between a ring containing a vortex and a reference BEC \cite{Corman2014qis}, both created through the Kibble-Zurek mechanism. 
f) Vortices resulting from artificial gauge fields \cite{Lin2009smf}. 
g) Pair of vortices \cite{neely2010observation} and turbulent vortex ensemble \cite{Kwon2014ros} generated by moving a barrier through the BEC at high speed.
h) Spiral interference signal between a vortexless and a 6-vortex phase pattern imprinted on an annular Fermi gas \cite{DelPace2022ipc}. 
Reproduced figures with permission from \cite{Donadello2014oos,Coddington2003oot,Corman2014qis,neely2010observation,Scherer2007vfb,Aidelsburger2017rdi,Kwon2014ros,Anderson2001wds,Ku2016cos,Madison2000vfi}. Copyright by the American Physical Society. Reproduced figures from \cite{Hadzibabic2006bkt,Lin2009smf}, with permission from Springer Nature.}
    \label{fig:methods}
\end{figure*}

\begin{itemize}

   \item \textit{Rotation.} The most intuitive and elegant way to introduce vorticity in a quantum gas, consists in making the gas rotate. This can be implemented by introducing an asymmetry in the trapping potential through off-center optical beams \mbox{\cite{Madison2000vfi,Abo-Shaeer2001oov,Wright2013dps}} or anisotropic magnetic confinement \mbox{\cite{Haljan2001dbe,Engels2002neo}} and having them rotate about the symmetry axis. For increasing rotation rates, more and more same-sign vortices enter the system and align their core lines to the rotation axis. The formation of regular triangular Abrikosov lattices, with a density proportional to the rotation rate, was observed in Refs. \mbox{\cite{Madison2000vfi,Engels2002neo, Abo-Shaeer2001oov}}. Such a vortex crystal can be perturbed by exciting Tkachenko modes \mbox{\cite{Coddington2003oot}} and can turn into a square lattice in the presence of two spin components in the trap \mbox{\cite{Schweikhard2004vld}}.
   
   \item \textit{Soliton decay.} Standard solitons are usually stable in one-dimensional systems, while they are subject to decay into other topological excitations in higher dimensions. The snake instability is a typical example of such a decay mechanism that shows how the solitonic plane (in 3D systems) or the soliton line (in 2D systems) breaks into vortex ring pairs or vortex dipoles. These vortical structures are, in fact, less energetic than solitons and hence more robust. The decay mechanism and the vortices produced were studied in 3D isotropic \cite{Anderson2001wds}, in elongated geometries \cite{MunozMateo2014csa,Donadello2014oos,Ku2016cos}, or in flat potentials \cite{Tamura2022oos}.
   
   \item \textit{Thermal activation in 2D systems.} Systems with reduced dimensionalities are characterized by enhanced fluctuations. In 2D, interactions drive a continuous phase transition from a superfluid (at low temperatures) to a normal gas (at high temperatures), called Berezinskii-Kosterlitz-Thouless transition. As the temperature is increased, the spatial correlations turn from algebraically to exponentially decaying and initially low-energy vortex-antivortex pairs unbind toward a proliferation of free vortices. This was observed in continuous systems \cite{Hadzibabic2006bkt,Choi2013oot} and in lattice configurations \cite{Schweikhard2007vpi}. 

   \item \textit{Merging isolated condensates.} When independent condensates are released from their traps and start to spatially overlap, they interfere, as demonstrates with pairs \cite{Andrews1997oii} or linear arrays of BECs \cite{Hadzibabic2004ioa}. Similarly, initially isolated condensates  with independent phases, arranged in on a ring, interfere and depending on their phase distributions, they can also show vortices in the final merged condensate. Such a mechanism was first demonstrated using three BECs on a ring \cite{Scherer2007vfb} and then extended to many \cite{Aidelsburger2017rdi}.
   
   \item \textit{Quenches across phase transitions.} 
   A phase transition is usually associated to a symmetry breaking.  When the critical point of the transition is rapidly crossed, it is likely to create domains with different phases in the system.    Through such a mechanism, turbulent phase patterns form, including the possibility to have vortices.   
   Spin vortices have been observed by quenching a spinor BEC across a quantum phase transition \cite{Sadler2006ssb} and also across the  normal to BEC phase transition \cite{Weiler2008svi}. This lead to an intense investigation of the Kibble-Zurek mechanism in quantum gases \cite{Lamporesi2013,Corman2014qis,Navon2015cdo,Chomaz2015eoc,Donadello2016cac,Goo2021dsi}.
   
   \item \textit{Artificial gauge fields.} Vortices can also be introduced through the application of synthetic magnetic fields. This possibility has been demonstrated \cite{Lin2009smf} by introducing a spin-orbit coupling potential made of a pair of Raman beams propagating along  different directions and having a frequency difference that matches an atomic two-level transition. These artificial fields do not possess the limitations of real rotating systems and can potentially lead to the observation of quantum Hall effects \cite{Dalibard2011cag}. Furthermore, due to the analogy of rotating neutral atoms and charged particles in a magnetic field, mechanical rotation at the trap frequency maps the Bose-Einstein condensate onto a single Landau gauge wavefunction, allowing access to the lowest Landau levels, with the rotating system behaving like a strongly correlated quantum Hall fluid \cite{Fletcher2021gsi,Mukherjee2022cob}.
   
   \item \textit{Strong mechanical perturbations.} By moving an obstacle such as a focused laser beam across the superfluid \cite{neely2010observation,Neely2013cot,Kwon2014ros,gauthier2019giant,Kwon2021sea}, by strongly shaking the confining potential \cite{henn2009emergence,Navon2016eoa}, or by letting two condensates collide with a large relative momentum \cite{Kim2017css}, the gas can be moved at velocities well above the local speed of sound, turning it into a turbulent system with strong phase fluctuations and many vortices. In this case, the net vorticity is on average zero. 
   
   \item \textit{Optical phase imprinting.} The spatial phase pattern characterizing vortices can be reproduced optically through digital micromirror devices or spatial light modulators. Illuminating the gas with such a pattern, the wavefunction acquires a local phase, proportional to the light intensity. This method was proposed in the past \cite{Denschlag2000gsb,Kumar2018psc}, but only recently implemented to imprint vortices in a controllable way \cite{DelPace2022ipc}. Optical imprinting techniques with dynamic beams configurations \cite{Matthews1999via} or static ones using Laguerre-Gauss beams \cite{Beattie2013pci} successfully produced quantum vortices.  

\end{itemize}

These methods have all been observed in gases with short-ranged contact interactions. In this work, we focus on the recent developments in the vortex generation in dipolar gases, in particular with the novel magnetostirring technique. We study the stability of the dipolar condensate for increasing rotation frequencies, then focus on the vortex core shape as a function of the relative angle between the rotation axis and the magnetic field, and finally extend from a single vortex to many, with the formation of exotic ordered vortex structures, typical of dipolar gases.

\section{Dipolar gases and magnetostirring}

Dipolar BECs comprised of highly-magnetic lanthanide atoms \cite{Lu2011sdb, Aikawa2012bec} are taking the lead as a promising platform to investigate complex quantum matter phenomena. Such systems can be viewed as the quantum version of classical ferrofluids, where the physics is dominated by the interplay of long-range anisotropic dipole-dipole interactions (DDIs) \cite{Norcia2021dia,Chomaz2022dpa} with short-range contact interactions and trap potentials. The balance of these contributions leads to excitations known as {\it rotons}. These low energy excitations occur at a local minimum in the dispersion relation. First predicted in liquid helium \cite{Landau1941hto}, they are considered as the precursor to crystallization due to their appearance at a finite momentum, and therefore of a fixed wavelength. Predicted in dipolar systems in 2003 \cite{ODell2003rig,Santos2003rms}, they were observed in dipolar gases first in cigar \cite{Chomaz2018oor,Petter2019ptr} then pancake \cite{Hertkorn2021dfa,Schmidt2021rei} geometries. The transition to a stable crystal was achieved later, where triggering a roton instability \cite{Chomaz2018oor} leads to an appearance of a density modulated state with global phase coherence, known as a supersolid \cite{Gross1957uto}, with translational symmetry broken along one \cite{Boettcher2019tsp,Tanzi2019ooa,Chomaz2019lla} or two \cite{Norcia2021tds,Bland2022tds} axes. Other observed phenomena in ultracold dipolar gases thus far include, but are not limited to, lattice spin models \cite{Lepoutre2019ooe,Patscheider2020cde}, extended Bose-Hubbard dynamics \cite{Baier2016ebh}, quantum droplets \cite{Chomaz2016qfd,FerrierBarbut2016ooq}, and the quantum Rosensweig instability \cite{Kadau2016otr}; constituting only the tip of the iceberg of possibilities during this second quantum revolution \cite{Norcia2021dia,Chomaz2022dpa}. Equivalent phenomena are also predicted in other systems with long-range interactions, such as ground-state heteronuclear molecules \cite{Carr2009cau,Schmidt2022sbd} and Rydberg atoms \cite{Adams2019raq}, and light-atom coupled systems \cite{Leonard2017sfi,Li2017asp,Goldman2014lig}.

In magnetic atoms, these phenomena arise due to the long-range nature of the interaction, which for a polarized gas of ultracold atoms presents as a pseudo-potential
\begin{align}
    U_\text{dd}(\textbf{r}) = \frac{C_\text{dd}}{4\pi}\frac{1-3\left(\hat{\mathbf{e}}\cdot\hat{\mathbf{r}}\right)^2}{r^3}\,,
    \label{eqn:DDI}
\end{align}
characterized by the dipole strength $C_\text{dd}=\mu_0\mu_m^2$ for atoms with a permanent dipole moment with magnetic moment $\mu_m$ and the vacuum permeability $\mu_0$, dipoles polarized along $\hat{\textbf{e}}$, and $\hat{\textbf{r}}=\textbf{r}/r$ with $r=|\textbf{r}|$.

\begin{figure}
    \includegraphics[width=0.6\columnwidth]{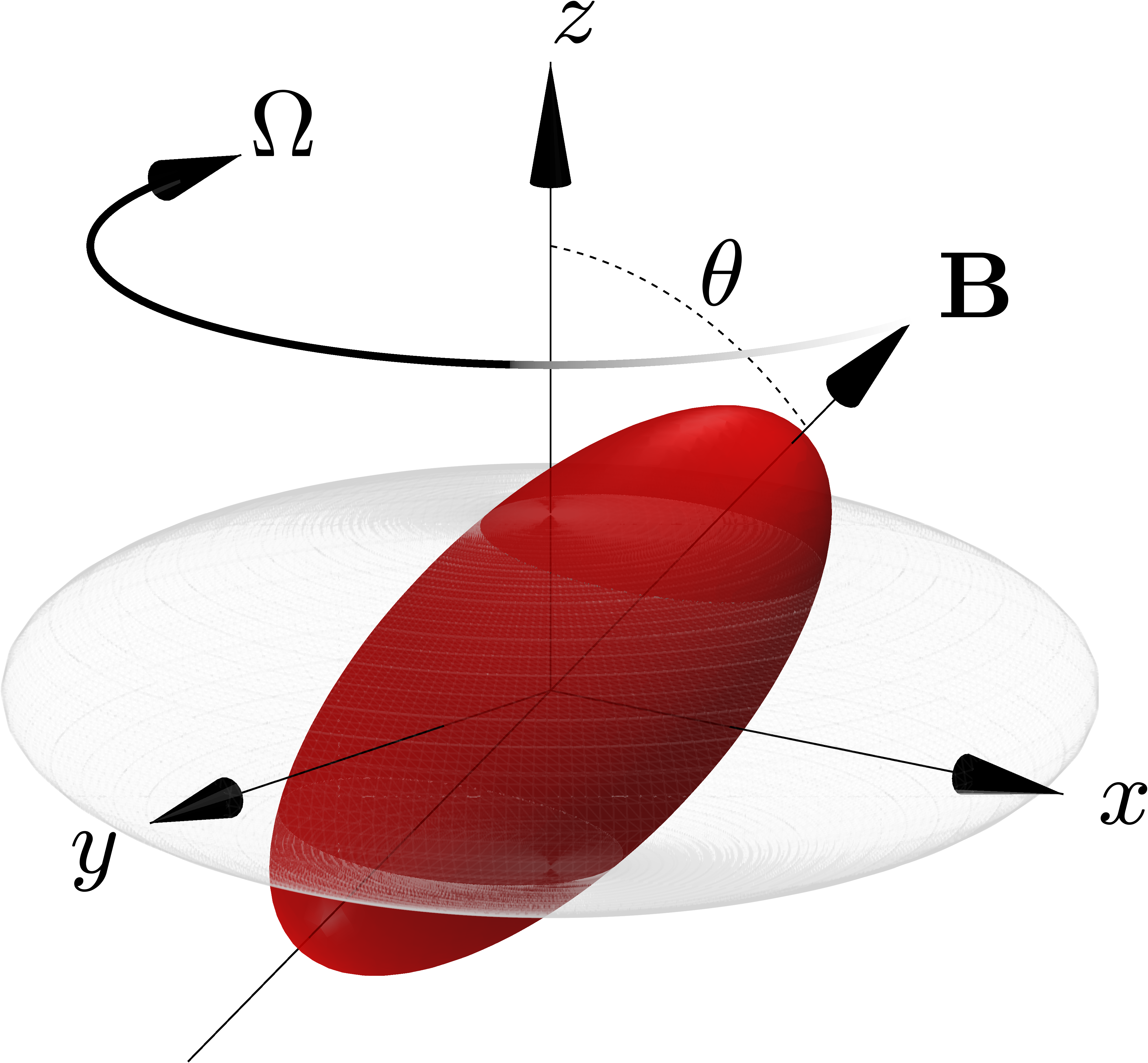}
    \caption{Illustration of magnetostirring. Dipolar BEC (red) in an azimuthally symmetric pancake trap (gray mesh) stretches along the B-field, at an angle $\theta$, and follows the polarization orientation as it is stirred at a frequency $\Omega$.}
    \label{fig:magnetostirring}
\end{figure}

The anisotropic nature of Eq.~\eqref{eqn:DDI} reveals a novel method of vortex generation. Here, two notions are important. First, the atomic magnets always align along the magnetic-field vector, which sets the polarization axis. Second, a polarized gas stretches along the polarization axis in a process known as magnetostriction \cite{ekreem2007aoo}, where dipoles align head-to-tail to maximize their attraction. This can be exploited to induce asymmetry into the dipolar BEC density profile, in an otherwise cylindrically symmetric trap, by tilting the magnetic field into the plane. Once tilted, the magnetic field can be rotated around the trap centre producing vortices in an analogous method to mechanical trap rotation of a non-dipolar BEC \cite{Prasad2019vlf}. See Fig.~\ref{fig:magnetostirring} for a schematic of this process. Recently, this method was employed to produce the first observed vortices in a dipolar gas \cite{klaus2022oov}, in which the vortices exhibit remarkable properties due to the dipole-dipole interaction, to be explored in the next section. Rapid magnetostirring is known to produce a different effect, where the atoms remain stationary but the dipole moments rapidly rotate \cite{Giovanazzi2002ttd}. This leads to a time-averaged dipole-dipole interaction that can be reversed from the bare DDI (i.e.~introducing a negative sign to Eq.~\eqref{eqn:DDI}), known as an anti-dipolar interaction, where head-to-tail anti-dipoles attract and side-by-side anti-dipoles repel. This was observed experimentally in 2018 \cite{Tang2018ttd}, though the condensate had a reduced lifetime due to destructive dynamical instabilities \cite{Prasad2019ior}. There remains hope that these can be suppressed by rotating at much larger frequencies \cite{Baillie2020rto}, and the impact of changing the trap geometry is unexplored. Furthermore, slow magnetostirring is also predicted to be a good probe of the stability of two-dimensional bright solitons \cite{tikhonenkov2008asi} and droplets \cite{li2023avq}, and may even produce vortices inside these intriguing self-bound objects \cite{Cidrim2018vis,Lee2018eoa}.

\section{Quantum vortices in dipolar gases}

The long-range and anisotropic DDI also drastically modifies the now well-established superfluid vortex properties \cite{Martin2017vav}. At the single vortex level, it is the anisotropy of the interaction that in turn produces elliptical vortex cores, observable when the vortex line and magnetic field vector are non-parallel \cite{yi2006vsi,abad2009vib,Ticknor2011asi,mulkerin2013anisotropic,mulkerin2014vortices}. Instead, under the condition of vortex lines parallel to the external magnetic field polarization, the presence of the roton minimum in the dispersion relation can produce concentric alternating rings of high and low density surrounding the vortex core \cite{yi2006vsi,Wilson2008mot,Ticknor2011asi,mulkerin2013anisotropic,mulkerin2014vortices,JonaLasinio2013rci}. These effects can be understood from a simple toy model introduced in a review by Martin {\it et al.}~\cite{Martin2017vav}: decomposing the condensate density for a vortex into $n(\textbf{r}) = n_0 - n_v(\textbf{r})$, with the vortex-free solution $n_0$ and the vortex density $n_v$, and calculating the dipolar energy contribution from this solution provides the necessary insight. The dipolar interaction energy is then given by
\begin{align}
2E_\text{dd} &= \int\text{d}^3\textbf{r}\int\text{d}^3\textbf{r}'\, n(\textbf{r}) U_\text{dd}(\textbf{r}-\textbf{r}') n(\textbf{r}') \nonumber\\
2E_\text{dd} &= \int\text{d}^3\textbf{r}\int\text{d}^3\textbf{r}'\, n_0 U_\text{dd}(\textbf{r}-\textbf{r}') n_0 \nonumber \\
&~+ \int\text{d}^3\textbf{r}\int\text{d}^3\textbf{r}'\, n_v(\textbf{r}) U_\text{dd}(\textbf{r}-\textbf{r}') n_v(\textbf{r}') \nonumber\\
&~- 2\int\text{d}^3\textbf{r}\int\text{d}^3\textbf{r}'\, n_0 U_\text{dd}(\textbf{r}-\textbf{r}') n_v(\textbf{r}')\,.
\label{eqn:dipE}
\end{align}
The negative sign of the last term in this expansion can be interpreted such that the vortex acts as a giant anti-dipole sitting inside the (normal) dipolar BEC. For dipoles aligned parallel to the vortex line, this means that the density is pulled towards the vortex core, causing the ripple structure. For dipoles tilted into the plane, the antidipolar interaction causes head-to-tail repulsion, hence the vortex core repels atoms along the B-field and stretches elliptically, analogously to magnetostriction. Anisotropic vortex cores have also been observed in vortices in superconductors due to the same effect \cite{Fuchs2022avs}.

The same line of thought elucidates the properties of vortex-vortex interactions. In a non-dipolar condensate confined to a quasi-2D geometry, vortex-vortex interactions are also long-ranged (scaling as $1/r$) but are isotropic, and the nature of the interaction (attractive/repulsive) is set by the sign of the circulation. Same sign vortices repel and rotate around one another at a fixed distance, whereas opposite sign vortices attract one another and annihilate. In 3D, there are also interesting recombination and rebounding effects between vortex filaments, depending also on the relative angle of approach and speed \cite{Serafini2017vra}. The isotropy of the vortex-vortex interaction leads to the triangular Abrikosov vortex lattice. In the dipolar system, we can see from Eq.~\eqref{eqn:dipE} that the inter-vortex interaction behaves like a bare dipole-dipole interaction. Two dipolar vortices sat in a side-by-side configuration (with respect to the B-field angle) will repel one another, and attract head-to-tail. Thus, there are predictions that there can be suppression of vortex-antivortex annihilation \cite{mulkerin2014vortices} and an elliptic precession of two same-sign vortices \cite{mulkerin2014vortices,gautam2014dynamics}. Beyond vortex pairs, long-range dipolar interactions also alter the bulk lattice structure. Dipoles pointing out of the plane--i.e.~interacting isotropically within the plane--give rise to a triangular lattice structure when the contact interaction is repulsive \cite{yi2006vsi,JonaLasinio2013rci}, but this can be square or rectangular for attractive or zero contact interactions \cite{Cooper2005vli,Zhang2005vli,Kumar2016tdv}. Breaking the isotropy by tilting the dipoles, breaks the symmetry of the lattice, where instead the vortices align in stripes, preferring to maximize head-to-tail configurations as much as possible \cite{yi2006vsi,cai2018vpa,Prasad2019vlf}. There are open questions pertaining to finite temperature effects on dipolar vortices, such as Kibble-Zurek physics and the Berezinskii-Kosterlitz-Thouless transition, though in the absence of external driving there is evidence of anisotropic and polarized turbulent decay \cite{Bland2018qft}. To our knowledge, the impact of quantum fluctuations on these predictions is unknown.

In spite of 20 years of theoretical and experimental endeavour, the first experimental evidence of vortices in a dipolar Bose-Einstein condensate was only observed in an experiment last year \cite{klaus2022oov}. Vortices in a dipolar BEC are predicted to be generated through all the aforementioned methods at the beginning of this article \cite{Martin2017vav}. However, Klaus \textit{et al.} utilized the magnetostirring protocol of Fig.~\ref{fig:magnetostirring}. Turbulent clouds of vortices were generated, and ordering into a stripe configuration was observed. This experiment was inspired by earlier theoretical works that predicted the formation of vortex lattices under continuous magnetostirring at magnetic fields tilted at an arbitrary angle from the trap geometry \cite{Prasad2019vlf,Prasad2021aar}. In future experiments, it will be interesting to see single vortex properties, and to generate ground state lattices under continuous rotation \cite{Prasad2019vlf}.

There are open questions on how quantum fluctuations change the predictions made thus far, though in the superfluid phase their effect is expected to be minimal. In the supersolid phase, however, they play the crucial role of stabilizing the gas against collapse, and the observation of vortices could provide an unambiguous smoking gun of superfluidity in supersolid states \cite{Roccuzzo2020ras,Gallemi2020qvi,Ancilotto2021vpi,sindik2022car,Gallemi2022spo}. In this special issue, we analyze the effect of varying the scattering length and dipolar angle on the predicted vortex properties--including single vortex properties and the resulting lattice formation--focussing on the experimentally relevant regime, and including beyond-mean-field effects.

\section{Theoretical model}

We focus our investigations on parameter regimes relevant to the recent experimental observation of vortices in a dipolar BEC \cite{klaus2022oov}. We use an extended Gross-Pitaevskii formalism for the wavefunction $\Psi\equiv\Psi(\textbf{r},t)$, where the time-dependent equation reads \cite{Waechtler2016qfi,FerrierBarbut2016ooq, Chomaz2016qfd,Bisset2016gsp}
\begin{align}
    i\hbar\frac{\partial\Psi}{\partial t} &= \bigg[-\frac{\hbar^2\nabla^2}{2m}+\frac12m\left[\omega_r^2(x^2+y^2)+\omega_z^2z^2\right] \nonumber \\ 
    &~+ \int\text{d}^3\textbf{r}'\, U(\textbf{r}-\textbf{r}')|\Psi(\textbf{r}',t)|^2  +\gamma_\text{QF}|\Psi|^3\bigg]\Psi(\textbf{r},t)\,,
    \label{eqn:GPE}
\end{align}
for atoms of particle mass $m$ confined in a harmonic trap with frequency $f_j = \omega_j/2\pi$ and aspect ratio $\gamma=\omega_z/\omega_r$,  reduced Planck's constant $\hbar$, and the wavefunction $\Psi$ is normalized to the total atom number $N=\int {\rm d}^3\mathbf{x}|\Psi|^2$. The last two terms of Eq.~\eqref{eqn:GPE} describes the interparticle interactions, determined by the pseudo-potential, 
\begin{align}
    U(\textbf{r}) = \frac{4\pi\hbar^2a_{\rm s}}{m}\delta(\textbf{r})+\frac{3\hbar^2a_\text{dd}}{m}\frac{1-3\left(\hat{\mathbf{e}}(t)\cdot\hat{\mathbf{r}}\right)^2}{r^3}\,,
\end{align}
where the first term derives from short-range interactions with strength given by the s-wave scattering length $a_s$. The second term is the anisotropic and long-ranged dipole-dipole interaction given by Eq.~\eqref{eqn:DDI}, characterized by dipole length $a_\text{dd}=C_\text{dd}m/12\pi\hbar^2$. We always consider $^{162}$Dy, such that $a_\text{dd}=129.2\,a_0$, where $a_0$ is the Bohr radius. The time dependent polarization $\hat{\mathbf{e}}(t)$ gives us the ability to model magnetostirring, with the dipoles polarized instantaneously along a time-dependent axis, given by
\begin{align}
    \hat{\mathbf{e}}(t)=(\sin\theta(t)\cos\phi(t),\sin\theta(t)\sin\phi(t),\cos\theta(t))\,,
\end{align}
with time dependent polarization angle $\theta(t)$ and $\phi(t)=\int_0^t \text{d}t'\Omega(t')$, for rotation frequency protocol $\Omega(t)$. Finally, beyond-mean-field effects are treated through the inclusion of a Lee–Huang–Yang correction term \cite{Lee1957eae,schutzhold2006mean,Lima2011qfi}
\begin{align}
    \gamma_\text{QF}=\frac{128\hbar^2}{3m}\sqrt{\pi a_s^5}\,\text{Re}\left\{ \mathcal{Q}_5(\edd) \right\} \, ,
\end{align}
with the auxiliary function $\mathcal{Q}_5(\edd)=\int_0^1 \text{d}u\,(1-\edd+3u^2\edd)^{5/2}$, and the relative dipole strength $\edd = a_{\rm dd}/a_{\rm s}$. We identify two regimes that are used throughout this work: the contact dominated regime $\edd<1$ and the dipole dominated regime $\edd>1$. The initial state $\Psi(\textbf{r},0)$ of the real-time simulations is obtained by first solving Eq.~\eqref{eqn:GPE} in imaginary time, giving $\Psi_0(\textbf{r})$, and then adding non-interacting noise to this state. This noise is generated through single-particle eigenstates $\phi_n$ and the complex Gaussian random variables $\alpha_n$ sampled with $\langle|\alpha_n|^2\rangle = (e^{\epsilon_n/k_BT}-1)^{-1}+\frac12$ for a temperature $T=20$ nK, such that the total initial state can be described as $\Psi(\textbf{r},0) = \Psi_0(\textbf{r}) + \sum_n' \alpha_n \phi_n(\textbf{r})$, where the sum is restricted only to the modes with $\epsilon_n\le2k_BT$ \cite{blakie2008das}. The simulations are performed with a split-step Fourier method, in a $192^3$ box of equal lengths $L=40\,\mu$m, imposing a spherical cut-off for the dipolar potential to avoid contamination of our results with phantom simulation copies \cite{Ronen2006bmo}.

\section{Stability of the condensate under magnetostirring}

The stability of a non-dipolar condensate under mechanical rotation is well understood from a hydrodynamic formalism \cite{Recati2001oov,Sinha2001dio}. This approach tells us that, in spite of the energetic favorability of generating vortices for rotation frequencies exceeding $\Omega\sim0.3\,\omega_r$, with symmetric radial trap frequency $\omega_r$, a dynamical instability is required to nucleate the vortices into the system. This phenomenon was first observed by the LKB group led by J. Dabilard in a non-dipolar BEC \cite{Madison2000vfi}. Typically, this dynamical instability occurs at $\Omega\sim0.7\,\omega_r$, where low-lying collective modes are seeded and become unstable \cite{Madison2000vfi,Hodby2001vni}--associated with a resonance with the quadrupole mode at $\Omega\sim\omega_r/\sqrt{2}$--providing the necessary avenue for vortex creation. 

This same phenomenon has been thoroughly investigated in dipolar gases, where in spite of the complexity of the underlying integro-differential equation there are analytic solutions in the Thomas-Fermi regime \cite{Eberlein2005eso,vanBijnen2010cef}. A key feature of the non-dipolar analysis is that there exists three solutions for the stationary elliptic parabolic density profile at large rotation frequencies: one stretched parallel to the trap deformation, and two stretched perpendicularly. Of the solutions aligned perpendicularly to the trap deformation, one is weakly deformed and highly stable, and the other is highly unstable and strongly deformed. In the dipolar system, under weak trap ellipticity and dipoles pointing out of the plane, all three solutions have been shown to exist, and possess similar features \cite{vanBijnen2007dio,martin2008iav,vanBijnen2009esa}. For stronger dipole-dipole interactions, both the critical value of $\Omega$ at which these three solutions exist can be reduced from the non-dipolar result, and the critical frequency of the dynamical instability is similarly decreased \cite{vanBijnen2009esa}. In these works, ellipticity is induced by breaking the radial trap symmetry, however when dipoles are pointing into the plane elliptic deformation is instead provided by magnetostriction, and angular momentum is induced through magnetostirring. Crucially, this method exhibits the same features from the mechanically rotating Thomas-Fermi analysis, following a series of earlier theoretical predictions from Prasad {\it et al.} \cite{Prasad2019ior,Prasad2019vlf}, work that was also extended to arbitrary rotation angles \cite{Prasad2021aar}. Rotation at an arbitrary angle is known to present unique features due to centrifugal forces pushing the condensate orientation towards the plane, tilted with respect to the trap or magnetic field angle, and can facilitate stable solutions for rotation frequencies greater than the radial frequency \cite{Prasad2021aar,prasad2020ssd}. 

\begin{figure}
    \includegraphics[width=1\columnwidth]{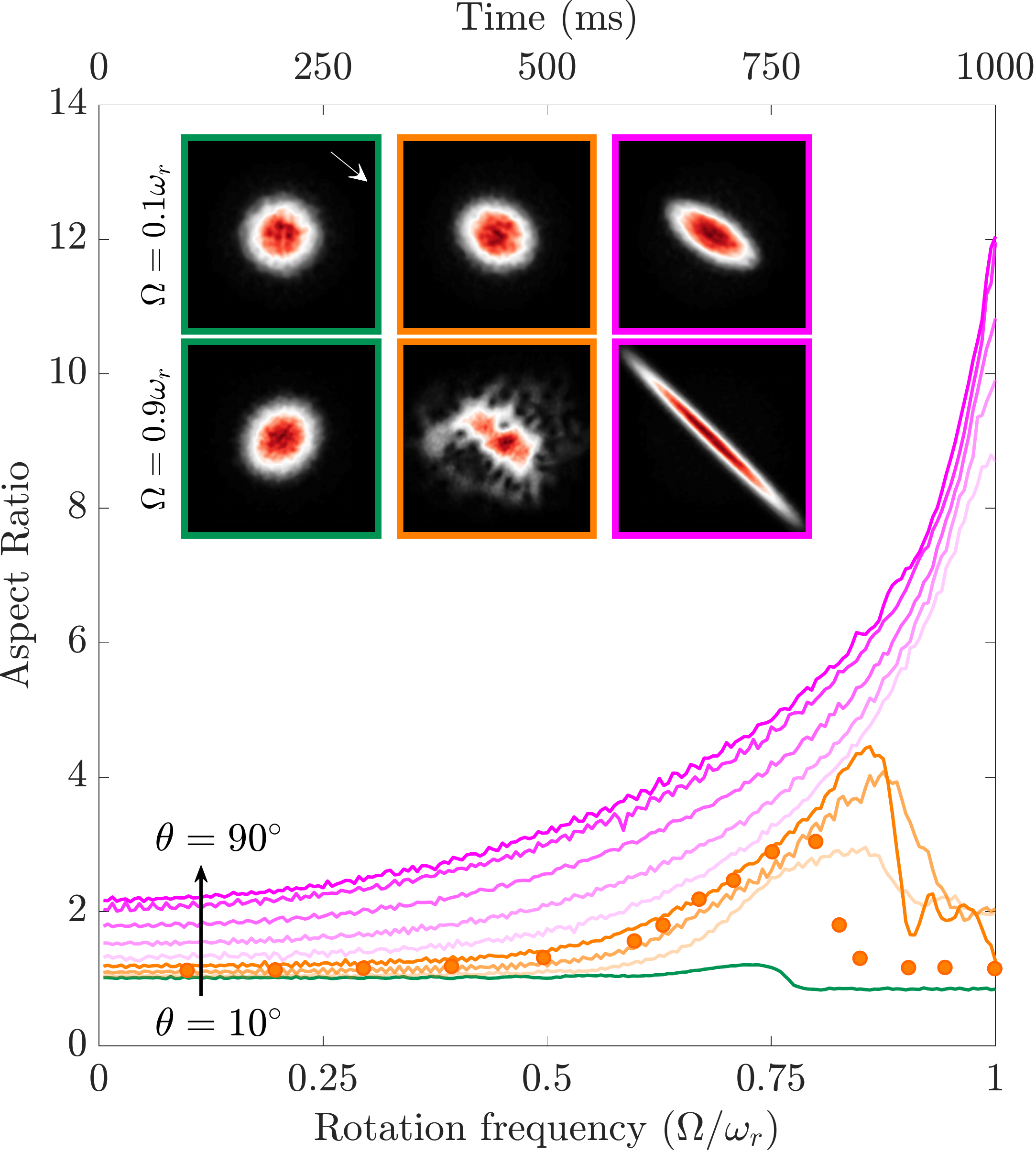}
    \caption{Aspect ratio of a $^{162}$Dy dipolar gas under accelerated magnetostirring. Each curve shows the aspect ratio evolution for increasing B-field tilt $\theta$, in steps of 10$^\circ$. Colors indicate which category the solutions fall under with (i) green, (ii) orange, and (iii) magenta (see text). The orange circles are the experimental data of Ref.~\cite{klaus2022oov} (Figure 1(c)) taken with the same system parameters at $\theta=35^\circ$. Insets show column densities for fixed time and increasing $\theta$, indicated by border color, and rotated to the same B-field angle given by the arrow. Each panel is $20\mu\text{m}\times20\mu\text{m}$. Parameters: $N=15000$, $a_s=110\,a_0$, $\vec{f}=(50,130)$ Hz, i.e.~$\gamma=2.6$.}
    \label{fig:stability}
\end{figure}

In a recent experiment, Klaus {\it et al.}~\cite{klaus2022oov} fixed the tilt angle of the magnetic field at $\theta=35^\circ$ and investigated the consequences of magnetostirring a dipolar BEC. In what follows, we investigate the significance of this choice of angle. We explore the stability of a dipolar BEC in the dominantly dipolar regime ($\epsilon_\text{dd}>1$) with fixed scattering length $a_s=110\,a_0$ under increased rotation frequencies, increasing at a rate of $2\pi\times 50$\,Hz s$^{-1}$, and different B-field tilt angles. The measured aspect ratio from the column density is shown in Fig.~\ref{fig:stability}. As expected, increasing the initial tilt angle the state is stretched due to magnetostriction, and we clearly observe the extension of the aspect ratio at low rotation frequencies. However, the stability of the state at fast rotations is angle-dependent and can be categorized into one of three groups: 
\begin{enumerate}[(i)]
    \item $0\le\theta<20^\circ$ (green): At shallow tilt angles the condensate density remains parabolic throughout the simulation, and although the aspect ratio initially increases, it finally decreases to $\sim0.85$, indicating that the stationary solution follows the middle branch from the Thomas-Fermi analysis. This branch is known to be highly stable, hence, no vortices are formed here. 
    \item $20^\circ\le\theta<50^\circ$ (orange): At intermediate angles the condensate density undergoes a dynamical instability (at $\Omega\sim0.8\,\omega_r$), and vortices begin to penetrate the condensate surface.
    \item $50^\circ\le\theta\le90^\circ$ (magenta): At deeper angles, as the condensate elongates, more atoms align in a head-to-tail configuration, and the attractive dipolar interaction holds the atoms together, suppressing a dynamical instability.
\end{enumerate}
The demarcating value of $\theta$ between each one of these categories will depend on the choice of scattering length and trap geometry. Remarkably, the solution for $\theta=90^\circ$ is a BEC for $\Omega=0$, but due to elongation of the state and increased head-to-tail attraction becomes a rotating trap-bound droplet state as $\Omega\to\omega_r$, stable due to the quantum fluctuations. We have verified that after another second of continuous driving at $\Omega=\omega_r$ the droplet continues to rotate without a growing collective mode amplitude. As the scattering length is increased, and the relative dipolar effects are reduced, the range of stable solutions for large $\theta$ (iii) will be decreased. Furthermore, the effect of magnetostriction will be diminished, such that increasing the scattering length plays a similar role as reducing the trap ellipticity in non-dipolar condensates, where this is known to aid stability. This, in turn, will also increase the $\theta$ range of stable solutions at small tilt angles (i). There, the ramp procedure is non-adiabatic compared to the timescales required to generate a dynamical instability, hence it is possible for the condensate to remain stable on the central stable branch predicted from Thomas-Fermi theory \cite{Prasad2019vlf,Prasad2021aar}. 

For this ramp procedure, however, there exists the Goldilocks range of angles (ii) where a dynamical instability occurs on a short timescale observable in experiments. After a collective mode goes unstable, vortices flood the condensate surface and the aspect ratio (obtained by a fit to a 2D Gaussian profile) no longer accurately captures the condensate density. Also overlaid on Fig.~\ref{fig:stability} are the data from Figure 1 of Ref.~\cite{klaus2022oov}, obtained with the same protocol but with $\theta = 35^\circ$ (orange points). At short times the data agree perfectly, it is only at late times, where the condensate undergoes a much earlier dynamical instability, that the results diverge. The time at which the instability happens will be sensitive to finite temperature effects and small inhomogeneities in the rotation procedure, which we attribute to the earlier onset of instability. The precise role of the trap aspect ratio $\gamma$ in the dipole dominated regime remains an open question, noting that previous studies have shown in the contact dominated regime that there exist trap aspect ratios where the dipolar interaction can fully suppress dynamical instabilities \cite{vanBijnen2010cef}.

\section{Single vortex in an oblate geometry}

Vortices in a dipolar BEC are known to have unique features compared to their non-dipolar counterparts \cite{Dalfovo2018ova}. For $\theta=0$ and scattering lengths close to a roton instability, the vortex core exhibits a high density ring around the core, and the core size tends to the inter-droplet distance as the scattering length is crossed into the supersolid phase. In tighter quasi-2D geometries, roton effects are more prominent prior to mean-field instability, and instead there are many concentric rings decreasing in amplitude radially. Tilting the B-field into the plane causes magnetostriction of the vortex core itself, and this has intriguing consequences on the resulting vortex lattice structure, that we will explore later.

\begin{figure}
    \includegraphics[width=1\columnwidth]{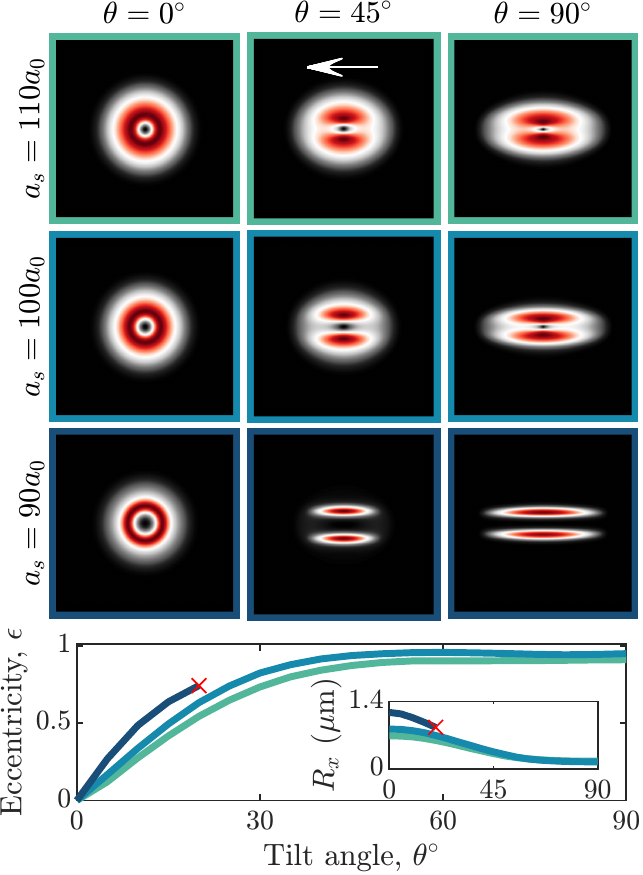}
    \caption{Column densities of a single vortex in a dipolar BEC, for varying tilt angle $\theta$ and scattering length $a_s$.
    For 90$a_0$ the condensate undergoes a roton instability at 25° (red crosses), and the stationary state is instead two tilted droplets. The core eccentricity, measured at an arbitrary low isodensity $3\times10^{19}$m$^{-3}$, tends to 1 for increasing $\theta$ in all cases. The inset shows the width of the minor axis. Other parameters the same as Fig.~\ref{fig:stability}.}
    \label{fig:vortex}
\end{figure}

For our system parameters we assess how the ellipticity and core size depend on both the scattering length and dipole angle, paying particular heed to the results in the range $20^\circ<\theta<50^\circ$, which we have seen can generate vortices within a reasonable experimentally feasible timescale ($<1\,$s). In Fig.~\ref{fig:vortex}, we show the column densities for a single vortex, for three tilt angles and three values of the scattering length. These stationary solutions are obtained through generating an initial condition with a 2$\pi$ phase winding in the $xy$-plane and evolving in imaginary time. For $\theta=0^\circ$, the core is cylindrically symmetric, increasing in radial size as the scattering length is lowered. At first glance, this contrasts the energy argument given around Eq.~\eqref{eqn:dipE}. However, as the scattering length is reduced the healing length increases, setting a wider vortex core size. In this trap geometry, the effect of the roton is present, but subtle, appearing as a slightly increased density around the core, and also widening the core size to be comparable to the roton wavelength \cite{Gallemi2020qvi}. For any non-zero tilt angle the vortex core is stretched into an ellipse. We measure the core eccentricity as $\epsilon=\sqrt{1-R_x^2/R_y^2}$, for major axis $R_y$ along the B-field and minor axis $R_x$ perpendicular to the B-field, taken at an arbitrary low isodensity $3\times10^{19}$\,m$^{-3}$. We see that the core eccentricity exceeds 0.5 for all considered scattering lengths when the tilt angle exceeds $\sim20^\circ$. Therefore, in the angle regime of interest ($20^\circ<\theta<50^\circ$) the effect of the dipolar interactions should be visible from the column density.

At deeper angles than 25$^\circ$, and small scattering lengths, the increased head-to-tail attraction triggers a roton instability, and the stationary state is a pair of droplets. Note that the ground state with these parameters is a single droplet with no vortex. Here, a vortex is imprinted between two droplets, though not visible in the density profile, highlighting the challenge of observing vortices in the density profile in the supersolid phase \cite{Roccuzzo2020ras,Gallemi2020qvi,Ancilotto2021vpi,sindik2022car,Gallemi2022spo}. One recent proposal \cite{sindik2022car} has suggested quenching the scattering length into the BEC phase after creating the vortex through rotation of the supersolid state. In this way, the vortex remains after the quench and would be visible in the density profile. One other method could be to create a persistent current in a ring supersolid, and observe the circulation through the interference pattern in time-of-flight imaging \cite{Tengstrand2021pci}.

\begin{figure*}
    \centering
    \includegraphics[width=2\columnwidth]{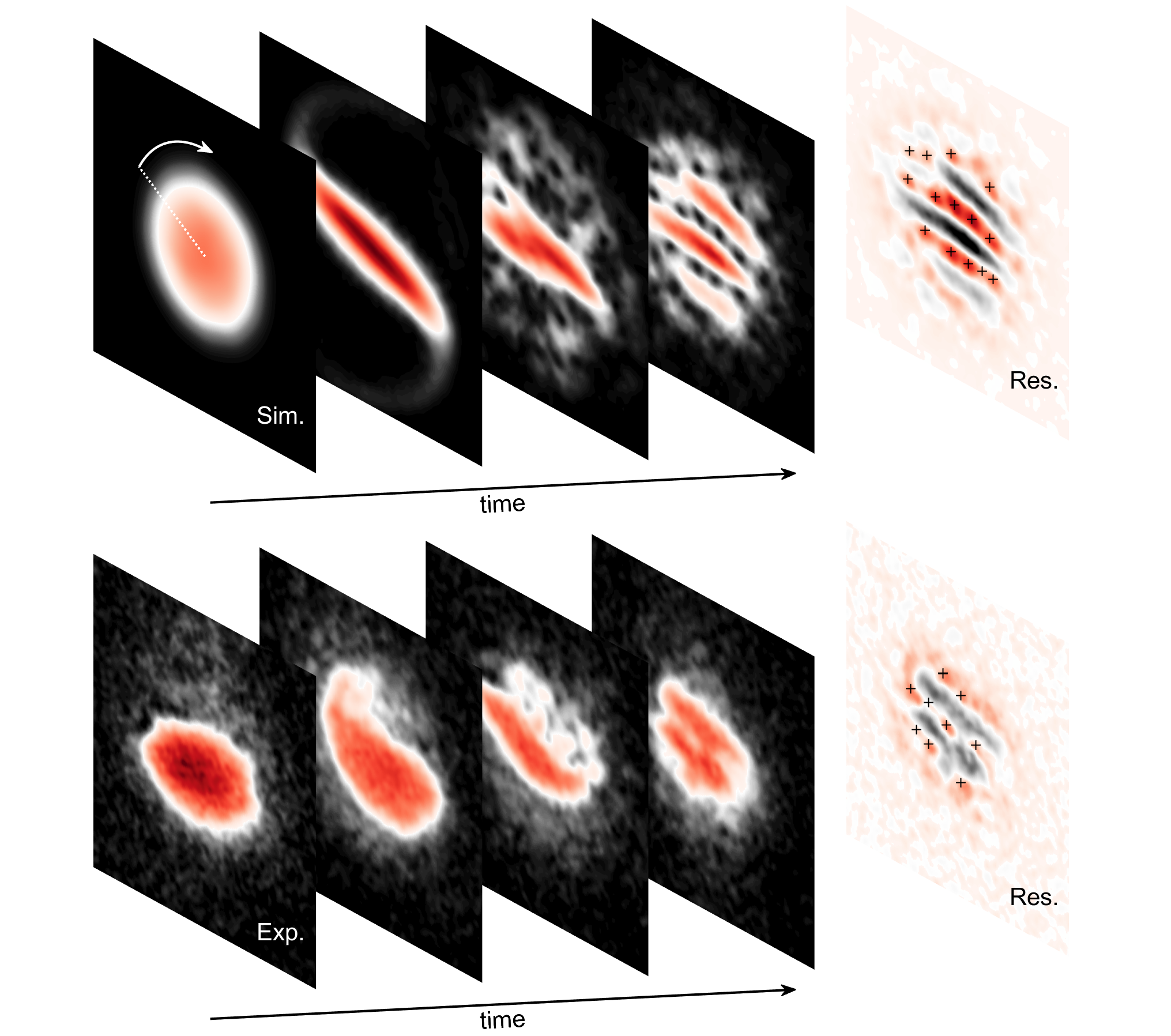}
    \caption{Generation of vortex stripes in a dipolar BEC. Top row: Time series of real-time simulation snapshots at (16, 164, 542, 1200)ms using the experimental parameters. Bottom row: Time series of the corresponding experimental snapshots at (20, 127, 314, 447)ms. For the last image of each series we applied the vortex detection algorithm~\cite{klaus2022oov} and show the calculated residuals, with the detected vortices marked with crosses. Used parameters: $\theta = 35^\circ$, $\Omega=0.7\,\omega_r$, $a_s = 109a_0$, $N=10000$, $\vec{f}=(50,130)$\,Hz.}
    \label{fig:latticeformation}
\end{figure*}

\section{From one to many vortices in a dipolar gas}

We have seen that the recipe to generate vortices through magnetostirring in a dipolar gas requires careful tuning of the dipolar parameters. Choosing an angle in the intermediate regime ($20^\circ<\theta<50^\circ$) is ideal for triggering the dynamical instability required to induce vortices. Moreover, individual vortices induced have an eccentricity $>$0.5, and dipolar effects are observable in the in situ density.

Rather than increasing $\Omega$ linearly from 0, as in Fig.~\ref{fig:stability}, previous experiments have investigated the formation of vortex lattices either through jumping to the target value of $\Omega\sim0.7\,\omega_r$ after generation of the BEC, or by cooling across the thermal cloud-to-BEC phase transition under continuous rotation at this value \cite{Haljan2001dbe}. Here, we continuously stir a dipolar BEC jumping from $\Omega=0$ to $\Omega=0.7\,\omega_r$ at $t=0$ with fixed $\theta=35^\circ$ to observe rapid dynamical instability and vortex lattice nucleation. In Fig.~\ref{fig:latticeformation} (top row) we show four simulation snapshots taken at; $t=0$, a static BEC; after 164 ms showing the formation of spiral arms, a hallmark feature of the growing dynamical instability; post instability at 542 ms with vortices flooding the condensate density; after 1200 ms of continuous rotation where the system is settling towards a static vortex stripe lattice. Vortex stripes---rows of vortices aligned along the magnetic field direction---were one of the first predicted consequences of the dipole-dipole interaction on vortex structure \cite{yi2006vsi}. We also show snapshots of the same process taken from the experimental data from Ref.~\cite{klaus2022oov} [Fig.~\ref{fig:latticeformation} (bottom row)]. Remarkably, similar features are observed, such as the appearance of spiral arms and the eventual production of vortices inside the density profile. In this experiment, due to atom losses, it was not possible to continuously stir until a lattice has formed, however vortex stripes were observed for the first time. These effects are readily observed in the residual images, where the average density from many experimental runs--which is smooth due to the random distribution of vortices--is subtracted from the single experimental image (see Ref.~\cite{klaus2022oov} for more details) leaving behind peaks at the vortex positions and troughs along the high density stripes. From our simulations, we have found that larger atom numbers facilitate a faster reorganization into the lattice, and there is hope that future experiments will be able to observe this exciting prediction.

\section{Vortex lattices}

\begin{figure}
    \includegraphics[width=1\columnwidth]{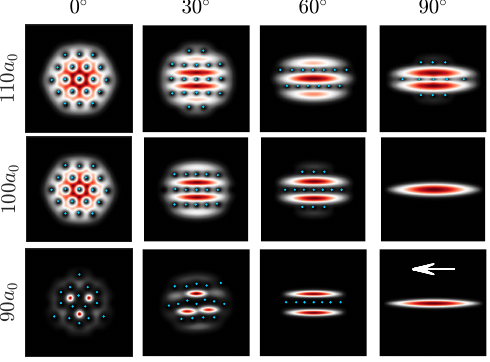}
    \caption{Vortex lattice solutions of a dipolar BEC. Light blue points mark vortex positions. Parameters: $\Omega = 0.7\omega_r$, other parameters are the same as Fig.~\ref{fig:stability}.}
    \label{fig:lattice}
\end{figure}

Finally, for our system parameters, we investigate the transition from triangular to stripe vortex lattice, and the fate of the vortices across the BEC-to-supersolid transition. 
In Fig.~\ref{fig:lattice}, we present the stationary state lattice solutions found through imaginary time propagation of Eq.~\eqref{eqn:GPE} with $\Omega=0.7\,\omega_r$. When $\theta=0^\circ$ the lattice is triangular, as in non-dipolar BECs. We note, however, that a square lattice here is only 0.03\% larger in energy, and we do not rule out that a square lattice would be possible in this geometry, similar to the results of Refs.~\cite{Cooper2005vli,Zhang2005vli,Kumar2016tdv} but with positive scattering length. Already at $\theta=30^\circ$ we find that the vortex lattice takes on the stripe structure, with the stripes more prominent at lower scattering lengths. This is consistent with the findings of Klaus \textit{et al.}, who observed vortex stripes at smaller scattering lengths (approx.~$105\,a_0$). At deeper angles still, the number of stripes is reduced as the magnetostriction stretches the condensate along the magnetic field direction, until the condensate becomes a single rotating droplet and the energy cost of adding a vortex exceeds that obtained from maintaining an irrotational flow. The critical rotation frequency to induce a vortex across the BEC-to-droplet transition is an open question, and would be an interesting future study.

Vortices in the supersolid regime are known to possess intriguing properties, taking on the core size of the inter-droplet spacing \cite{Gallemi2020qvi}, and snaking between the droplets in order to interact with one another \cite{Ancilotto2021vpi}. Typically, the crystal structure of two-dimensional supersolids are known to be triangular lattices, with some evidence of the possibility of square lattices \cite{young2022sls}, and predictions for exotic labyrinthine-like and honeycomb crystal solutions at high densities \cite{Zhang2019saa,Zhang2021pos,Hertkorn2021pfi}, resembling the nuclear pasta states of neutron stars. The energy minimum position of a vortex is in the centre of each triangle, however there is a metastable minimum between each droplet pair \cite{Ancilotto2021vpi}, and vortices fill all of these minima forming the supersolid vortex lattice at fast rotation frequencies \cite{Gallemi2020qvi}. Rotating high density honeycomb structure supersolids is predicted to break apart the crystal \cite{Gallemi2022spo}. In our system, the stationary state with $a_s = 90a_0$ is a supersolid for all angles and $\Omega=0.7\,\omega_r$, however is an unmodulated BEC at $\Omega = 0$, hinting at the intriguing possibility of triggering the unmodulated-to-modulated transition through rotation. This behaviour has been predicted previously \cite{Tengstrand2021pci}, due to the two competing avenues for angular momentum production in a supersolid: either a vortex can enter contributing to the superfluid angular momentum, a contribution proportional to the superfluid fraction $f_s$, or the classical angular momentum $\Omega m\langle x^2+y^2\rangle$ can be increased by reducing the superfluid fraction, i.e. inducing supersolidity. Intermediate tilt angles display a complex interplay between maintaining a fixed interdroplet spacing and the vortices attempting to align in stripes, until at deeper tilt angles where the droplet number reduces to one droplet without vortices.

\section{Conclusion and outlook}

In conclusion, we have shown that rotation at intermediate tilt angles are ideal for generating vortices in a short timescale, delicately balancing the increase of ellipticity with deeper tilt angles against the increased head-to-tail attraction suppressing the necessary dynamical instability. We have shown that the predictions for the shape and size of dipolar vortices in the absence of beyond-mean-field corrections remain present in the dominantly dipolar regime, where these corrections are necessary for stability. Whereas previous studies, that have neglected beyond-mean-field effects, required $\theta\approx70^\circ$ in order to see the transition from triangular to square lattice \cite{cai2018vpa}, we are able to access the dominantly dipolar regime where this transition occurs as low as $30^\circ$. Recent experimental results only present the first steps to observe a large quantity of intriguing properties. At finite temperatures,  the impact of the dipolar interaction on universal scaling laws for Kibble-Zurek physics and the nature of the BKT transition in dipolar gases remains unknown. Future research directions may investigate the properties of dipolar vortices in binary condensates, such as dipolar spin mixtures between two spin states of a single species, multispecies mixtures such as erbium-dysprosium \cite{Trautmann2018dqm}, or dipolar non-dipolar mixtures such as dysprosium-potassium \cite{Ravensbergen2018poa} and erbium-ytterbium \cite{schafer2023realization}. Predictions thus far have investigated the interplay between roton immiscibility--the two species separating at the roton wavelength--and vortex lattices \cite{zhang2016exotic,kumar2017vortex,kumar2018vortex,kumar2019spatial,tomio2020dipolar}. Recent theoretical investigations into binary systems have revealed novel interweaving supersolid states \cite{li2022long,bland2022alternating} and miscible supersolid states \cite{scheiermann2023catalyzation}. In liquid helium, a second component was utilized in order to directly image the vortices from the bulk component \cite{bewley2006visualization}, and the same ideas may be brought to binary supersolidity as an ideal platform to image vortices in supersolids, for example extending the work of Ref.~\cite{Anderson2000vpi} to dipolar gases. Prospects for rotating an unmodulated BEC into the supersolid regime \cite{Tengstrand2021pci}, and tests of superfluidity of the supersolid, are currently a hotly debated topic in the field \cite{Tanzi2021eos,norcia2022cao,Roccuzzo2022moi}.

\section*{Acknowledgments}

In December of 2021, Jean Dalibard was awarded with the most prestigious French prize for physicists, the CNRS Gold medal. This paper is a contribution to celebrate his numerous achievements throughout his prestigious scientific carrier. In the last 40 years, Professor Dalibard enormously contributed to the field of ultracold gases manipulation in several fundamental ways, always finding the key ingredient to solve problems and elegant solutions to describe complex phenomena in a beautiful and intuitive manner. His research on laser cooling and invention of the magneto-optical trap was fundamental for the achievement of extremely low temperatures towards the successful realization of Bose-Einstein condensates \cite{Griffin1995bec}. He then played a crucial role in the investigation of the many-body properties of polarized condensates and spinor systems, from collective modes to topological excitations.
Undoubtedly, one of his major and fundamental contributions is related to vortex physics, with his experiments on the first rotating condensates and the observation of vortex lattices, with his studies on quasi two-dimensional gases in harmonic and flat potentials, from the observation of vortices created via the Kibble-Zurek mechanism \cite{Kibble1976toc,Zurek1985cei} to the characterization of the Berezinskii-Kosterlitz-Thouless transition \cite{Berezinskii1972dol,Kosterlitz1973oma}.

We are grateful to E. Poli and the Innsbruck erbium-dysprosium team for helpful discussions. This study received support from the European Research Council through the Advanced Grant DyMETEr (No. 1010545), the QuantERA grant MAQS by the Austrian Science Fund FWF (No. I4391-N), the DFG/FWF via Dipolare E2 (No. I4317-N36) and a joint-project grant from the FWF (No. I4426). Part of the computational results presented have been achieved using the HPC infrastructure LEO of the University of Innsbruck. G. L. acknowledges financial support from Provincia Autonoma di Trento. T.B. acknowledges financial support by the ESQ Discovery programme (Erwin Schrödinger Center for Quantum Science \& Technology), hosted by the Austrian Academy of Sciences (ÖAW).

\end{document}